\documentclass[]{ceurart} 

\usepackage{booktabs}
\usepackage{multirow}
\usepackage{graphicx}
\usepackage{xcolor}
\usepackage{amsmath}
\usepackage[normalem]{ulem}
\usepackage{arydshln}

\newcommand{\sid}{\textsc{SID}}
\newcommand{\tit}{\textsc{Title}}
\newcommand{\realdelta}{\Delta}

\definecolor{reasoninggain}{HTML}{16803A}
\definecolor{reasoningloss}{HTML}{B42318}
\newcommand{\gain}[1]{\textcolor{reasoninggain}{\bfseries #1}}
\newcommand{\loss}[1]{\textcolor{reasoningloss}{\bfseries #1}}

\definecolor{revision}{HTML}{000000}
\newcommand{\revise}[1]{#1}

\begin{document}

\copyrightyear{2026}
\copyrightclause{Copyright for this paper by its authors.
  Use permitted under Creative Commons License Attribution 4.0
  International (CC BY 4.0).}

\conference{GenAI-eCom: Agentic and Generative AI for E-Commerce Workshop at RecSys 2026, September 2026, Minneapolis, Minnesota, USA}


\title{The Disconnect Between Better Descriptive Reasoning Trace Quality and Recommendation Effectiveness}




\author[1]{Gustavo Penha}[%
  email=gustavop@spotify.com,
]
\address[1]{Spotify, United States}

\author[2]{Juan Elenter}[%
  email=juane@spotify.com,
]
\address[2]{Spotify, United States}

\author[3]{Claudia Hauff}[%
  email=claudiah@spotify.com,
]
\address[3]{Spotify, Netherlands}

\author[4]{Hugues Bouchard}[%
  email=hb@spotify.com,
]
\address[4]{Spotify, Spain}

\author[5]{Paul Bennett}[%
  email=pbennett@spotify.com,
]
\address[5]{Spotify, United States}

\author[6]{Mounia Lalmas}[
  email=mounial@spotify.com
]
\address[6]{Spotify, United Kingdom}

\begin{abstract}
Recent work has focused on improving explicit natural-language descriptive reasoning traces for generative recommendation. This includes systems that augment semantic ID (SID) prediction with chain-of-thought reasoning. However, because SIDs are opaque learned identifiers rather than natural language, they require costly alignment before an LLM can reason over them. 
This provides a controlled experimental setting in which both item representation (Title vs.\ SID) and semantic grounding (minimal vs. extensive SID alignment) can be varied independently.
We therefore present the first controlled comparison of descriptive reasoning trace quality across semantic IDs and natural-language titles in a 2 × 2 factorial study on three Amazon product domains using a shared Qwen3-1.7B backbone. We find that introducing explicit descriptive reasoning traces reduces traditional offline recommendation effectiveness under standard SFT and RL training, even though natural language titles produce substantially more grounded and interpretable traces.  Extensive SID alignment improves descriptive trace quality but not traditional offline recommendation effectiveness, while a richer reward signal partially recovers performance.
Overall, our results show that improving descriptive reasoning trace quality is not, by itself, sufficient to consistently improve traditional offline recommendation effectiveness under the training objectives and evaluation protocols studied here.

\end{abstract}


\maketitle


\section{Introduction}


Generative recommendation reformulates item prediction as autoregressive token generation: given a user's interaction history, a language model decodes the multi-token identifier of the next item token-by-token~\cite{Rajput2024TIGER}. 
The dominant approach represents items as semantic IDs (SIDs)---discrete identifiers learned from item content using residual quantization. SIDs have evolved from research prototypes into production systems serving millions of users~\cite{DAmico2026GLIDE}, and recent work has demonstrated that SID-based LLMs can unify search, recommendation, and reasoning over catalogs containing more than 10 million items~\cite{DeNadai2026NEO}. As a result, SIDs have become the foundation for a new generation of reasoning-based recommender systems.


With SIDs established as a viable production technology, recent research has shifted from improving item representations to augmenting recommendation models with explicit natural-language reasoning traces. These traces describe user histories, inferred preferences, and recommendation rationales before predicting the next item. Multi-stage training, reinforcement learning, and alignment techniques have substantially improved their quality, establishing reasoning-augmented generative recommendation as an emerging research direction~\cite{He2026SIDReasoner,Liu2025OneRecThink,Zhao2026HoloRec,Guo2026PROMISE}.


However, generating descriptive reasoning traces over SIDs requires substantially more training than reasoning over natural language. Because SID tokens are opaque learned identifiers rather than part of an LLM's pretrained vocabulary, the model must first learn what these identifiers represent before it can generate meaningful traces over them. We refer to this additional effort as the \emph{alignment tax}: the training stages, auxiliary objectives, and model capacity required to bridge the gap between opaque item identifiers and the LLM's native language understanding. Existing systems illustrate this cost. SIDReasoner dedicates an entire alignment stage with eight auxiliary objectives~\cite{He2026SIDReasoner}, OneReason identifies insufficient semantic grounding as a key limitation of earlier approaches~\cite{OneReason2026}, and recent work shows that when such grounding is inadequate, reasoning can even reduce recommendation accuracy by up to 25\%~\cite{Zhang2026WhyThinkingHurts}.


Natural-language item representations offer an alternative that avoids aligning opaque identifiers to an LLM's language space. 
Because an LLM already understands words such as ``strategy game'' or ``wireless headphones,'' it can, in principle, reason directly over item descriptions without first learning the semantics of newly introduced identifier tokens. By contrast, introducing hundreds of opaque SID tokens may compete with the language capabilities acquired during pretraining~\cite{Kirkpatrick2017EWC}, including those underpinning chain-of-thought reasoning. 
This raises a broader question: \emph{if natural-language titles and extensive SID alignment both improve  reasoning trace quality, do those improvements translate into better recommendation effectiveness?}

We present the first controlled study of how item representation influences the descriptive quality of explicit reasoning traces in generative recommendation. Using a shared Qwen3-1.7B backbone, identical data splits, and shared teacher-generated traces, we compare semantic IDs and natural-language titles with and without explicit reasoning across three Amazon product domains. Our central finding is that substantial improvements in reasoning trace quality do not consistently translate into improvements in traditional offline recommendation effectiveness.\footnote{By \emph{traditional offline evaluation} we mean user-based train/test splits where the model is evaluated against a single held-out interaction per user. Alternative offline evaluation paradigms include Cranfield-style relevance collections with pooled judgments~\cite{Penha2025LLMJudge,Smucker2025MovieLens} and LLM-as-a-judge assessments~\cite{Fabbri2025LLMJudgePodcast}, both of which can provide richer relevance signals than a single ground-truth item.}  Natural-language titles and extensive SID alignment both improve trace quality, yet neither consistently improves recommendation effectiveness under the training objectives studied here. By contrast, richer optimization objectives partially recover offline performance, highlighting the importance of optimization in translating higher-quality reasoning traces into improved recommendation effectiveness.

Our analysis provides insight into this disconnect. Reasoning over semantic IDs produces generic, poorly grounded traces, whereas reasoning over natural-language titles produces substantially more grounded and interpretable traces. Yet these improvements in trace quality do not translate into better recommendation effectiveness. Even extensive SID alignment improves trace quality without improving ranking effectiveness, while optimizing with an LLM judge that evaluates both trace quality and recommendation relevance partially recovers recommendation performance. 
Together, these findings suggest that item representation strongly influences the quality of generated reasoning traces, but higher-quality traces alone do not consistently translate into better traditional offline recommendation effectiveness.

Our contributions are threefold:
\begin{itemize}
    \item We present the first controlled comparison of descriptive reasoning traces across semantic IDs and natural-language titles, isolating the effects of item representation and semantic grounding in a shared experimental framework.
    \item We show that two independent interventions that substantially improve descriptive reasoning trace quality---natural-language titles and extensive SID alignment---are not sufficient to  consistently improve traditional offline recommendation effectiveness under the training objectives studied here.
    \item We show that richer reinforcement learning objectives partially recover traditional offline recommendation effectiveness, indicating that optimization plays a central role alongside descriptive reasoning traces.
\end{itemize}


\section{Related Work}


\paragraph{Generative Recommendation with Semantic IDs.}
TIGER~\cite{Rajput2024TIGER} introduced the paradigm of representing items as discrete semantic IDs learned with RQ-VAE and predicting them autoregressively. Subsequent work has improved semantic ID construction~\cite{Guo2025HiDVAE}, addressed cold-start challenges~\cite{ColdStart2026}, investigated shared search and recommendation spaces~\cite{10.1145/3705328.3759300}, and studied scaling behavior~\cite{Ju2025SIDHandbook}. These studies focus on improving semantic IDs as representations for recommendation. Our work complements this line of research by studying how item representation influences descriptive reasoning trace quality.

\paragraph{Reasoning in Generative Recommendation.}
Chain-of-thought reasoning has rapidly emerged as a major research direction in generative recommendation.
Most retrieval-oriented approaches reason over semantic IDs.
SIDReasoner~\cite{He2026SIDReasoner} introduces multi-stage alignment followed by reinforcement learning (RL), OneRec-Think~\cite{Liu2025OneRecThink} combines supervised fine-tuning with RL,
HoloRec~\cite{Zhao2026HoloRec} interleaves reasoning with hierarchical encoding,
and PROMISE~\cite{Guo2026PROMISE} uses process reward models to guide reasoning at inference.
More recent work explores semantic-guided latent reasoning (S2GR~\cite{S2GR2026}),
step-aligned policy optimization (SAPO~\cite{SAPO2026}),
and reasoning with verification (VRec~\cite{Zhu2026VRec}).
Ranking-oriented methods such as GR2~\cite{GR22026},
ReRec~\cite{ReRec2026}, R2Rec~\cite{R2Rec2025}, and Reasoning to Rank~\cite{Zheng2026ReasoningToRank}
reason over hybrid or natural-language item representations,
while RecZero~\cite{Kong2025RecZero} and GREAM~\cite{Wang2025GREAM}
extend reasoning to other recommendation settings.
Despite this growing body of work, existing studies evaluate explicit reasoning within a single item representation. Our work complements this literature by providing the first controlled comparison of descriptive reasoning traces across semantic IDs and natural-language titles using a shared experimental framework.

\paragraph{The Cost of Making Reasoning Work with SIDs.}
A recurring pattern in this literature is the substantial engineering effort required to make reasoning effective over opaque semantic IDs. Existing approaches introduce increasingly complex mechanisms to bridge the gap between learned identifiers and an LLM's language understanding. SIDReasoner~\cite{He2026SIDReasoner} relies on a dedicated alignment stage with eight multi-task objectives, while OneReason~\cite{OneReason2026} argues that successful reasoning additionally requires semantic grounding (``perception'') and interest abstraction (``cognition''). Other methods introduce semantic supervision~\cite{S2GR2026}, step-aligned reinforcement learning~\cite{SAPO2026}, or verification during reasoning~\cite{Zhu2026VRec}. Although these approaches differ in implementation, they all incur additional training stages, auxiliary objectives, or specialized optimization to enable reasoning over semantic IDs.


When these alignment requirements are not met, reasoning over semantic IDs can even degrade recommendation quality. Zhang et al.~\cite{Zhang2026WhyThinkingHurts} diagnose a ``linguistic inertia'' effect, where chain-of-thought shifts attention from collaborative SID evidence toward natural-language context, reducing recommendation accuracy by up to 25\%. TwiSTAR~\cite{TwiSTAR2026} similarly finds that reasoning is beneficial only for difficult examples, motivating adaptive strategies that selectively invoke chain-of-thought. 
Our work complements these efforts by asking whether this additional engineering translates into improvements in both descriptive reasoning traces and recommendation effectiveness.


\paragraph{Latent Reasoning.} An alternative line of work bypasses natural-language reasoning traces entirely. ReaRec~\cite{ReaRec2025} feeds hidden states back through the recommender with reasoning position embeddings, LARES~\cite{LARES2025} uses a depth-recurrent architecture with flexible test-time compute scaling, and LatentR3~\cite{LatentR32025} reasons over compact latent tokens with efficiency comparable to no-reasoning baselines. Together, these approaches suggest that some benefits of reasoning can be obtained without generating explicit natural-language traces. In contrast, our work studies reasoning through natural-language traces and asks how the choice of item representation influences their effectiveness.


\paragraph{Test-Time Compute Scaling.} Recent theoretical and empirical work explains why extended reasoning can improve model performance. Additional reasoning tokens increase effective computational depth~\cite{Feng2023CoTTheory,Merrill2024CoTExpressiveness}, and even semantically meaningless ``pause tokens'' can improve accuracy by providing additional forward passes~\cite{Goyal2024PauseTokens}. Empirically, compute-optimal test-time allocation~\cite{Snell2024TestTimeCompute}, reinforcement learning for reasoning~\cite{DeepSeekR12025}, and extended chain-of-thought in recommendation~\cite{Liu2025InferenceScaling} all demonstrate the benefits of increased reasoning computation.  
Our work complements this literature by studying how item representation influences explicit descriptive reasoning traces and whether improvements in those traces translate into recommendation effectiveness. In particular, we show that reasoning over opaque semantic IDs behaves differently from reasoning over natural-language item representations.
However, Kambhampati et al.~\cite{Kambhampati2025StopAnthropomorphizing} caution against interpreting intermediate tokens as genuine reasoning, arguing that they function as learned prompt augmentations whose task performance is largely independent of their semantic content. Our findings are consistent with this view: improvements in the descriptive quality of intermediate traces do not consistently translate into improved recommendation effectiveness.

\section{Experimental Framework}



We conduct a controlled 2×2 factorial study to determine whether improvements in descriptive reasoning traces translate into improved recommendation effectiveness. The two experimental factors are (1) item representation (SID vs. Title) and (2) explicit reasoning (No Reasoning vs. Reasoning), yielding four model configurations trained under an otherwise shared pipeline. This section describes the task, item representations, reasoning integration, the shared training pipeline, and our primary evaluation metric. Reasoning models are evaluated after supervised fine-tuning (SFT) and reinforcement learning (GRPO).

The factorial design allows us to independently vary item representation and explicit reasoning while keeping the underlying recommendation task, model architecture, and supervision fixed, thereby isolating the contribution of each factor.


\subsection{Problem Setup}


We formulate generative recommendation as next-item prediction. Given a user interaction history $\mathcal{H}_u = [i_1, i_2, \ldots, i_t]$, the task is to predict the next item $i_{t+1}$. A generative recommender models $P(i_{t+1}\mid\mathcal{H}_u)$ by autoregressively generating an item identifier $\text{id}(i_{t+1})$, represented either as a semantic ID (\sid) or its  natural-language title (\tit).

\subsection{Item Representations}

One experimental factor is how items are represented. We compare two representations of the same item catalog:

\paragraph{Semantic IDs (\sid).}
We use the released RQ-VAE semantic IDs from SIDReasoner~\cite{He2026SIDReasoner}: three-level residual quantization with a codebook of size 256 per level, trained on E5-large embeddings of item titles and descriptions. Each item is represented by a three-token code, e.g., \texttt{<s0\_42><s1\_17><s2\_203>}. These $3\times256=768$ special tokens are added to the LLM vocabulary.

\paragraph{Titles (\tit).}
Each item is represented by its raw title, truncated to 32 tokens. During inference, generated titles are resolved to catalog items using BM25 retrieval over the item index.

\subsection{Reasoning Integration}

For each representation, we train comparable models with and without chain-of-thought reasoning.


\paragraph{No reasoning.}
The model is fine-tuned to predict the next item's identifier directly from the interaction history.
Training examples use the format:
\texttt{The user interacted with [id($i_1$)], [id($i_2$)], \ldots\ Predict the next item:}.


\paragraph{With reasoning.}
We reuse the released reasoning traces from SIDReasoner~\cite{He2026SIDReasoner}, avoiding the need to regenerate teacher reasoning traces for approximately 49K training sequences per domain. Each trace consists of a natural-language explanation of why $i_{t+1}$ follows from $\mathcal{H}_u$, referring to item attributes, categories, and user preference patterns.

To isolate the effect of item representation on reasoning traces, we derive two representation-specific versions
of each trace by replacing item references with the corresponding identifier format:
\begin{itemize}
    \item \textbf{\sid:} item mentions are replaced with SID token sequences (e.g., ``\textit{the strategy RPG}'' $\rightarrow$ \texttt{<s0\_42><s1\_17><s2\_203>}).
    \item \textbf{\tit:} item mentions are replaced with the corresponding item titles.
\end{itemize}


This ensures that the reasoning content is identical across both configurations; only the item representation within the trace and the prediction target differ. Models are fine-tuned using the format
\texttt{[history] <think> [reasoning trace] </think> [id($i_{t+1}$)]}. Figure~\ref{fig:gt-trace-example} illustrates reasoning traces of both representations.

\begin{figure}[t]
\centering
\small
\fbox{\parbox{0.95\columnwidth}{
\textbf{History:} Paper Mate InkJoy Ballpoint Pens $\mid$ AmazonBasics Permanent Markers $\mid$ AmazonBasics Stapler\\[4pt]
\textbf{Target:} Mead Organizher Weekly Planner\\[6pt]
\textbf{\sid{} ground-truth trace:}\\
\texttt{<think>} ``I observe that the user is consistently engaging with practical office supplies characterized by utility and reliability, as seen in their selections: \texttt{<a\_255><b\_228><c\_102>}, \texttt{<a\_166><b\_189><c\_113>}, and \texttt{<a\_204><b\_62><c\_235>}. The preference for bulk packs and essential tools suggests an emphasis on productivity and readiness for daily tasks\ldots'' \texttt{</think>}\\
\texttt{<a\_212><b\_83><c\_6>}\\[4pt]
\textbf{\tit{} ground-truth trace:}\\
\texttt{<think>} ``I observe that the user is consistently engaging with practical office supplies characterized by utility and reliability, as seen in their selections: Paper Mate InkJoy 100ST Ballpoint Pens, Medium Point, Black, Box of 12 (1951257), AmazonBasics Permanent Markers, Black, 12-Pack, and AmazonBasics Stapler with 1000 Staples - Black. The preference for bulk packs and essential tools suggests an emphasis on productivity and readiness for daily tasks\ldots'' \texttt{</think>}\\
Mead Organizher My Week Canvas Planner (38827)
}}
\caption{Ground-truth training example from SIDReasoner for both representations (Office Products). A GPT-generated teacher trace explains why the target follows from the history. In the \sid{} variant, items are referenced by opaque SID codes; in the \tit{} variant, by their catalog titles. The reasoning narrative is otherwise identical. Thus the reasoning content is held constant while only the item representation differs.
}
\label{fig:gt-trace-example}
\end{figure}

\subsection{Training Procedure}


Both representations are trained using the same three-stage pipeline, following SIDReasoner~\cite{He2026SIDReasoner} with several modifications to enable a controlled comparison between semantic IDs and titles.\footnote{Compared with SIDReasoner, we: (1) use LoRA rather than full fine-tuning for Stage~1 to obtain lightweight no-reasoning baselines; (2) train separate no-reasoning models instead of evaluating reasoning-trained models in ``non-thinking mode'', allowing the reasoning delta to isolate the effect of adding reasoning; and (3) mix 50\% reasoning and 50\% no-reasoning samples during Stage~2 with a $20\times$ loss upweight on item-ID tokens to preserve the base prediction task.}

\begin{figure*}[t]
\centering
\includegraphics[width=\textwidth]{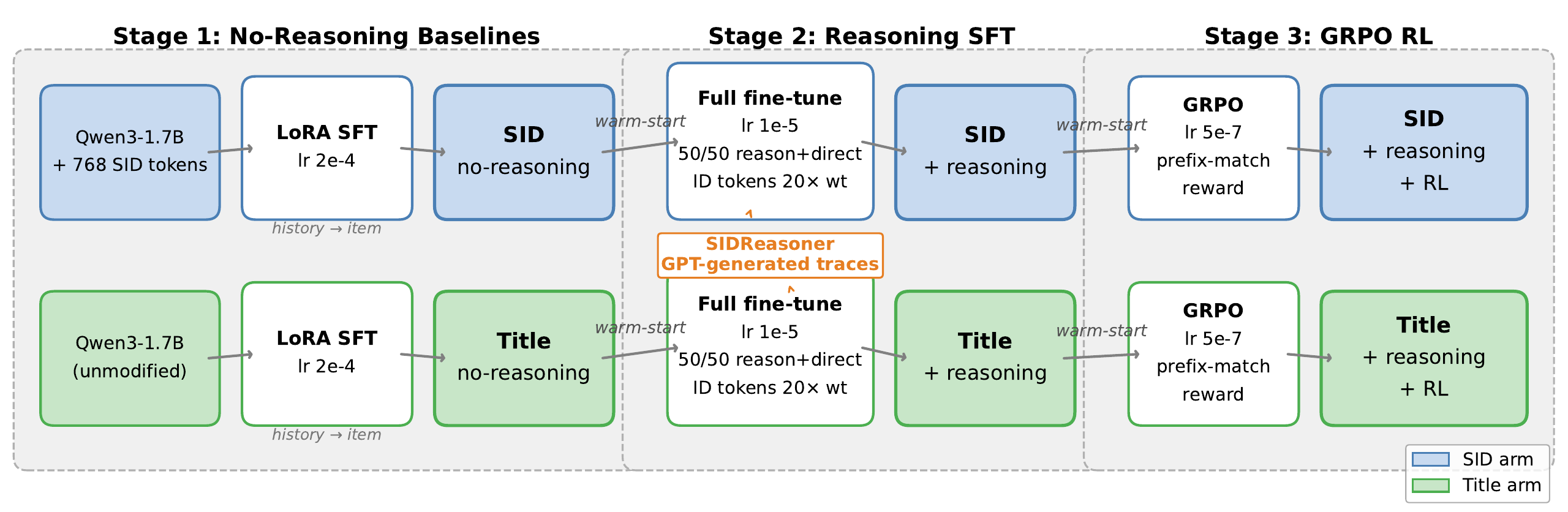}
\caption{Three-stage training pipeline for the $2 \times 2$ factorial study. Stage~1 trains no-reasoning baselines with LoRA; Stage~2 warm-starts from each baseline and adds reasoning via full fine-tuning on GPT-generated traces; Stage~3 applies GRPO reinforcement learning with a prefix-match reward.}
\label{fig:training-pipeline}
\end{figure*}

Figure~\ref{fig:training-pipeline} summarizes the overall training pipeline. Unless otherwise stated, all models use Qwen3-1.7B with early stopping on validation Recall@10 (patience 3). The SID configuration adds 768 special tokens to the vocabulary, whereas the Title configuration uses the original vocabulary with left truncation (\texttt{max\_seq}=512) to preserve the prediction target in long interaction histories.

\paragraph{Stage~1: No-Reasoning Baselines.}
Stage~1 trains a no-reasoning baseline for each representation using direct history-to-item prediction. Models are trained with LoRA (rank 16, $\alpha=32$), learning rate $2\times10^{-4}$, and effective batch size 128.

\paragraph{Stage~2: Reasoning SFT.}
Stage~2 teaches the model to generate a reasoning trace before predicting the next item. Models are warm-started from the corresponding Stage~1 checkpoint and fully fine-tuned (learning rate $10^{-5}$) using SIDReasoner's released GPT-generated reasoning traces.

Each training batch contains 50\% reasoning examples and 50\% direct history-to-item examples. This mixture preserves the underlying recommendation task while learning reasoning. For reasoning examples, item-ID tokens receive a $20\times$ larger loss weight than reasoning tokens to prioritize accurate item prediction.

\paragraph{Stage~3: GRPO RL.}
Stage~3 optimizes recommendation quality using Group Relative Policy Optimization (GRPO), starting from the Stage~2 checkpoint. For each prompt, we sample 16 completions and compute group-relative advantages across them.

We use ordered prefix matching between the predicted and target identifiers: for SIDs, matching the first $k$ of three hierarchical codes yields reward $k/3$; for titles, matching the first $k$ whitespace tokens yields reward $k/n$. Because the reward depends only on the predicted item, policy gradients are applied only to item tokens by masking the reasoning tokens. This preserves the reasoning behavior learned during Stage 2 while allowing RL to optimize recommendation quality.

\subsection{The Key Metric: Reasoning Delta \label{sec:reasoning-delta}}


For each representation $r \in \{\sid,\tit\}$, we define the \emph{reasoning delta} as the change in recommendation effectiveness due to adding chain-of-thought reasoning:
\begin{equation}
    \realdelta_r =
    \text{Metric}(r,\text{reasoning})
    -
    \text{Metric}(r,\text{no reasoning}).
\end{equation}

If the alignment tax is the primary bottleneck for reasoning over semantic IDs, we would expect
$\realdelta_{\tit} > \realdelta_{\sid}$: natural-language titles, which require no identifier alignment, should benefit more from reasoning than semantic IDs. The hypothesis follows from the intuition that LLMs can directly reason over natural-language item descriptions, whereas reasoning over semantic IDs first requires learning the semantics of opaque identifier tokens. We test this hypothesis in the following sections.


\section{Experimental Setup}

We evaluate the controlled comparison described in Section~3 on publicly available benchmarks using a shared evaluation protocol for both item representations.

\subsection{Datasets}
We use SIDReasoner's released Amazon Reviews 2018 splits~\cite{He2026SIDReasoner} to ensure direct comparability with prior work. All datasets are 5-core filtered (retaining only users and items with at least five interactions) and use chronological 8:1:1 train/validation/test splits with sliding-window histories of maximum length 10.

\begin{table}[t]
\caption{Dataset statistics after preprocessing.}
\label{tab:datasets}
\centering
\small
\begin{tabular}{lccc}
\toprule
\textbf{Dataset} & \textbf{Users} & \textbf{Items} & \textbf{Interactions} \\
\midrule
Amazon Video Games  & 9,053  & 3,858 & 61,417 \\
Amazon Office       & 7,638  & 3,459 & 48,656 \\
Amazon Industrial   & 6,842  & 3,686 & 45,325 \\
\bottomrule
\end{tabular}
\end{table}

\subsection{Evaluation Protocol}


We evaluate both representations on the full item catalog (i.e., no sampled negatives). The \sid{} configuration uses beam search with trie-constrained decoding, whereas the \tit{} configuration generates unconstrained text that is subsequently mapped to catalog items via BM25 retrieval to enable an item-level comparison rather than an exact string-generation evaluation. \revise{This decoding asymmetry is a potential confound when interpreting the \tit{} arm: reasoning-induced changes in generation style could degrade BM25 matching independently of recommendation quality. To quantify this effect, we measure the title resolution rate---the fraction of generated titles that faithfully reproduce a catalog entry---with and without reasoning (Appendix~\ref{app:bm25-resolution}).}


All metrics are computed at the \emph{item level} (i.e., comparing target and predicted ASINs---Amazon's unique product identifiers), rather than at the identifier level. Because semantic IDs are produced by vector quantization, different items can occasionally share the same identifier. When multiple items share an identifier, a predicted SID is expanded into all matching items at consecutive rank positions (e.g., a collision at rank~1 with three items occupies ranks~1--3). A hit therefore requires the correct \emph{item}, preventing collisions from artificially inflating retrieval metrics.

In practice, SID collisions are rare ($<1\%$), but they can still occur because RQ-VAE quantizes continuous embeddings to discrete codebook entries, causing similar items to share the same code. We therefore evaluate all models at the item level.


We report Recall@$k$ and nDCG@$k$ for $k\in\{5,10\}$. Statistical significance is assessed using paired bootstrap resampling (1,000 iterations, $\alpha=0.05$). Each test instance consists of predicting the chronologically last interaction in the user's held-out sequence. Unless otherwise noted, all reported recommendation metrics refer to traditional offline next-item prediction metrics computed at the item level.

This protocol ensures that differences between \sid{} and \tit{} reflect representation and reasoning behaviour rather than artifacts of the evaluation procedure.


\section{Results}


We organize the results around four questions: (1) How does reasoning affect recommendation effectiveness across item representations? (Section~5.1) (2) Does paying the full alignment tax improve reasoning? (Section~5.1.1) (3) Can an LLM-judge reward recover any degradation? (Section~5.1.2) (4) Does better reasoning translate into better recommendations? (Section~5.2).

\subsection{Main Results: The \texorpdfstring{$2 \times 2$}{2x2} Comparison}

%
%
We begin by asking whether introducing explicit reasoning improves traditional offline recommendation effectiveness across the two item representations. Table \ref{tab:main} reports the complete 2×2 comparison between \sid{} and \tit{} with and without explicit reasoning.
Unless otherwise stated, all results use the item-level evaluation protocol described in Section~4. Reasoning results correspond to the Stage~2 (SFT) and Stage~3 (GRPO) models introduced in Section~3. SASRec~\cite{Kang2018SASRec} is included as a non-generative baseline.

\begin{table*}[t]
\caption{Main results across three Amazon domains. Stage~2 = SFT on GPT-generated traces (full fine-tune); Stage~3 = GRPO RL with prefix-match reward. $\realdelta$: delta vs.\ no-reasoning baseline. Best no-reasoning per metric in \textbf{bold}. \revise{$^\dagger$: significantly different from no-reasoning baseline ($p < 0.05$, paired bootstrap, 1{,}000 iterations).} $^\star$Reference numbers from~\citet{He2026SIDReasoner} under same data splits, full fine-tuning, and SID-level evaluation (not item-level as in our rows).}
\label{tab:main}
\centering
\resizebox{\textwidth}{!}{%
\begin{tabular}{ll cccc cccc cccc}
\toprule
& & \multicolumn{4}{c}{\textbf{Video Games}} & \multicolumn{4}{c}{\textbf{Office Products}} & \multicolumn{4}{c}{\textbf{Industrial \& Scientific}} \\
\cmidrule(lr){3-6} \cmidrule(lr){7-10} \cmidrule(lr){11-14}
\textbf{Repr.} & \textbf{Stage} & R@5 & R@10 & N@5 & N@10 & R@5 & R@10 & N@5 & N@10 & R@5 & R@10 & N@5 & N@10 \\
\midrule
\multirow{5}{*}{\sid}
& 1: No reasoning & .0497 & .0728 & \textbf{.0356} & .0430 & \textbf{.1360} & \textbf{.1595} & \textbf{.1123} & \textbf{.1198} & .1013 & \textbf{.1361} & .0791 & \textbf{.0902} \\
& 2: Reasoning SFT & .0454 & .0689 & .0309 & .0384 & .1334 & .1589 & .1056 & .1138 & .1013 & .1324 & .0761 & .0862 \\
& 3: GRPO RL & .0451 & .0682 & .0314 & .0388 & .1317 & .1591 & .1037 & .1125 & .1008 & .1332 & .0761 & .0866 \\
\cdashline{2-14}
& $\realdelta$ SFT\% & $-$9\% & $-$5\% & $-$13\% & $-$11\% & $-$2\% & $-$0\% & $-$6\% & $-$5\% & 0\% & $-$3\% & $-$4\% & $-$4\% \\
& $\realdelta$ RL\% & $-$9\% & $-$6\% & $-$12\% & $-$10\% & $-$3\% & $-$0\% & $-$8\% & $-$6\% & $-$0\% & $-$2\% & $-$4\% & $-$4\% \\
\midrule
\multirow{5}{*}{\tit}
& 1: No reasoning & \textbf{.0510} & \textbf{.0783} & .0354 & \textbf{.0443} & .1350 & .1587 & .1107 & .1185 & \textbf{.1094} & .1253 & \textbf{.0800} & .0852 \\
& 2: Reasoning SFT & \revise{.0412$^\dagger$} & \revise{.0627$^\dagger$} & \revise{.0274$^\dagger$} & \revise{.0343$^\dagger$} & \revise{.1192$^\dagger$} & \revise{.1416$^\dagger$} & \revise{.0927$^\dagger$} & \revise{.1001$^\dagger$} & .1059 & .1240 & .0765 & .0825 \\
& 3: GRPO RL & \revise{.0391$^\dagger$} & \revise{.0614$^\dagger$} & \revise{.0269$^\dagger$} & \revise{.0341$^\dagger$} & \revise{.1188$^\dagger$} & \revise{.1397$^\dagger$} & \revise{.0915$^\dagger$} & \revise{.0985$^\dagger$} & .1032 & .1233 & .0744 & .0811 \\
\cdashline{2-14}
& $\realdelta$ SFT\% & $-$19\% & $-$20\% & $-$23\% & $-$23\% & $-$12\% & $-$11\% & $-$16\% & $-$16\% & $-$3\% & $-$1\% & $-$4\% & $-$3\% \\
& $\realdelta$ RL\% & $-$23\% & $-$22\% & $-$24\% & $-$23\% & $-$12\% & $-$12\% & $-$17\% & $-$17\% & $-$6\% & $-$2\% & $-$7\% & $-$5\% \\
\midrule
\multicolumn{14}{l}{\textit{Reference numbers from~\citet{He2026SIDReasoner}$^\star$}} \\
\midrule
---     & SASRec$^\star$       & .0489 & .0535 & .0293 & .0293 & .0860 & .0952 & .0664 & .0741 & .0724 & .0847 & .0523 & .0598 \\
\multirow{2}{*}{\sid}
        & TIGER$^\star$        & .0489 & .0763 & .0300 & .0402 & .1270 & .1429 & .1037 & .1121 & .1003 & .1325 & .0823 & .0924 \\
        & SIDReasoner$^\star$  & .0710 & .1031 & .0460 & .0563 & .1373 & .1648 & .1119 & .1208 & .1109 & .1438 & .0905 & .1010 \\
\bottomrule
\end{tabular}
}
\end{table*}

\subsubsection{Recommendation Effectiveness Across Representations}
Section \ref{sec:reasoning-delta} hypothesized that if the alignment tax is the primary bottleneck to effective reasoning, natural-language titles should benefit more from explicit reasoning than semantic IDs ($\realdelta_{\tit} > \realdelta_{\sid}$). Table \ref{tab:main} does not support this prediction. Across all three datasets, introducing explicit reasoning reduces traditional offline recommendation effectiveness for both representations, with negative or zero reasoning deltas in every setting. Surprisingly, the degradation is consistently smaller for \sid{} than for \tit{}. \revise{No \sid{} degradation reaches statistical significance ($p > 0.05$, paired bootstrap), whereas the \tit{} degradations on Video Games and Office are significant ($p < 0.05$; $\dagger$ in Table~\ref{tab:main}). The statistically reliable penalty is thus specific to \tit{} on two of three domains.} This finding is the opposite of what the alignment-tax hypothesis predicts and motivates a closer examination of whether insufficient semantic grounding is the underlying cause.

\revise{Because the \tit{} arm generates unconstrained text resolved via BM25, the larger reasoning penalty could partly reflect reasoning-induced drift in generation style that degrades BM25 matching, rather than changes in the model's item preferences. Appendix~\ref{app:bm25-resolution} analyzes this confound directly on the raw generations. Reasoning does not reduce the fraction of generations that are faithful catalog-title text on any domain (e.g.\ $98.0\%$ for both arms on Office), and an oracle resolution bound shows that even perfect resolution of the small drifted residue ($2$--$4\%$ of generations) would not close the gap. The degradation therefore reflects changes in the model's item predictions rather than a string-matching artifact.}

Stage~3 GRPO, which directly optimizes recommendation quality through policy gradients, does not recover this degradation. Across all \sid{} configurations, GRPO remains within $\pm0.001$ R@10 of its SFT warm-start, with $\realdelta_{\sid}$ values comparable to those after SFT. Likewise, for \tit{} on Video Games, GRPO yields a $\realdelta$ of $-$22\% R@10, essentially unchanged from the $-$20\% SFT penalty. Across all datasets, the best validation checkpoints appear early (100--400 training steps) before performance plateaus, suggesting that changing the optimization objective alone is insufficient to recover the degradation introduced by the current reasoning formulation.

To understand why GRPO fails, we analyze the diversity and reward distribution of generated completions. For each prompt, we sample 16 completions (matching the GRPO training configuration) from both the SFT and GRPO checkpoints and score them with the training reward. Surprisingly, GRPO generations receive \emph{lower} mean reward than their SFT counterparts (e.g., $0.001$ vs.\ $0.044$ on Games/Title), indicating that policy optimization drifts away from the SFT solution rather than improving upon it.

The underlying cause is reward sparsity. Between 70\% and 96\% of prompts produce 16 generations that all receive zero reward, yielding zero advantage and therefore no learning signal. On Games/Title, where only 7.8\% of SFT generations receive any partial match, reasoning collapses almost entirely: mean trace length falls from 69.6 words (SFT) to 1.1 words (GRPO) as the model converges to empty \texttt{<think>} blocks. Industrial/Title avoids this collapse because its higher partial-match rate (25.5\%) provides sufficient gradient to sustain reasoning. One possible explanation is that our lightweight Stage 1 provides insufficient semantic grounding for reasoning over semantic IDs. We test this hypothesis next.

\subsubsection{Paying the Alignment Tax}


A natural explanation for the degradation observed in Section~5.1.1 is that our lightweight Stage~1 provides insufficient semantic grounding. Unlike our setup, SIDReasoner~\cite{He2026SIDReasoner} precedes reasoning with an eight-task alignment phase designed to ground semantic IDs in natural language. If the alignment tax simply reflects insufficient alignment, then paying it in full should recover the benefits of reasoning.


To test this hypothesis, we reproduce SIDReasoner's complete Stage~1 for the \sid{} configuration on all three domains, denoted \sid{}-A (Table~\ref{tab:grpo-judge}). The alignment stage combines eight objectives covering next-item prediction, bidirectional title--SID translation, cross-representation recommendation, item and sequence grounding, and general reasoning. We then apply the same Stage~2 (reasoning SFT) and Stage~3 (GRPO) pipeline used throughout the rest of the paper.


Paying the alignment tax does not improve the effect of reasoning on traditional offline recommendation effectiveness (Table~\ref{tab:grpo-judge}, \sid{}-A rows). The eight-task alignment produces a stronger no-reasoning model than our lightweight baseline on two of the three datasets (Office R@10: $.1609$ vs.\ $.1595$; Industrial: $.1416$ vs.\ $.1361$), confirming that the additional alignment successfully improves the underlying representation. On Video Games, the aligned baseline is slightly weaker ($.0624$ vs.\ $.0728$), but remains competitive.


However, once reasoning is introduced, performance deteriorates substantially. Relative to its aligned no-reasoning baseline, Stage~2 loses 17--23\% R@10 across the three domains, and GRPO further degrades performance. Reasoning on the fully aligned model is consistently worse than reasoning on our lightweight baseline, with R@10 dropping to $.0519$, $.1239$, and $.1134$ on Games, Office, and Industrial, compared with $.0689$, $.1589$, and $.1324$ for the corresponding lean configurations.

These results isolate the source of the degradation. Because the fully aligned baseline is competitive with—and on most datasets stronger than—our lightweight baseline, the failure cannot be attributed to semantic IDs that the model does not understand. 
Improving semantic grounding strengthens the underlying recommendation model, but does not improve the benefit obtained from explicit reasoning. As we show in Section~\ref{sec:trace-quality}, this same pattern extends to descriptive reasoning traces themselves: better descriptive reasoning traces do not necessarily translate into improved traditional offline recommendation effectiveness.

\subsubsection{Recovering Performance with an LLM-Judge Reward}
\label{sec:grpo-judge}

The analysis above suggests that the prefix-match reward is too sparse to guide effective learning. Between 70--96\% of prompts yield zero reward across all 16 sampled generations, leaving the model with almost no learning signal. Moreover, the reward provides no credit to recommendations that are relevant but differ from the single held-out ground-truth item~\cite{Penha2025LLMJudge}. We therefore replace the accuracy-only reward with a composite objective that combines recommendation accuracy with LLM-judge assessments of reasoning quality and recommendation relevance:
\begin{equation}
    R = 0.5 \cdot R_{\text{acc}} + 0.25 \cdot R_{\text{trace}} + 0.25 \cdot R_{\text{rel}}.
\end{equation}
Here, $R_{\text{acc}}$ is the original prefix-match reward, $R_{\text{trace}}$ scores the quality of the reasoning trace (groundedness and coherence), and $R_{\text{rel}}$ scores recommendation relevance (Appendix~\ref{app:rl-reward-prompt}). Both judge scores are computed by GPT-4o-mini for each sampled completion. Unlike the accuracy-only variant, GRPO+Judge applies policy gradients to both reasoning and item tokens so that the model can optimize the reasoning behaviours evaluated by the judge. Table~\ref{tab:grpo-judge} summarizes the resulting performance.

\begin{table}[t]
\caption{R@10 across representations and RL variants. GRPO (acc): prefix-match reward only; GRPO+Judge: composite LLM-judge reward. \textbf{Bold}: highest R@10 per representation per domain. $^\dagger$: significantly different from no-reasoning baseline ($p < 0.05$, paired $t$-test). \sid{}-A: SIDReasoner's eight-task alignment Stage~1.}
\label{tab:grpo-judge}
\centering
\small
\begin{tabular}{lllllll}
\toprule
\textbf{Repr.} & \textbf{Stage} & \textbf{Games} & \textbf{Office} & \textbf{Industrial} \\
\midrule
\multirow{4}{*}{\sid}
& 1: No reasoning       & \textbf{.0728} & .1595 & .1361 \\
& 2: Reasoning SFT      & .0689 & .1589 & .1324 \\
& 3: GRPO (acc)          & .0682 & .1591 & .1332 \\
& 3: GRPO+Judge          & .0705 & \textbf{.1646} & \textbf{.1401} \\
\midrule
\multirow{4}{*}{\sid{}-A}
& 1: No reasoning       & \textbf{.0624} & \textbf{.1609} & \textbf{.1416} \\
& 2: Reasoning SFT      & .0519$^\dagger$ & .1239$^\dagger$ & .1134$^\dagger$ \\
& 3: GRPO (acc)          & .0344$^\dagger$ & .1200$^\dagger$ & .1165$^\dagger$ \\
& 3: GRPO+Judge          & .0361$^\dagger$ & .1254$^\dagger$ & .1169$^\dagger$ \\
\midrule
\multirow{4}{*}{\tit}
& 1: No reasoning       & \textbf{.0783} & \textbf{.1587} & .1253 \\
& 2: Reasoning SFT      & .0627$^\dagger$ & .1416$^\dagger$ & .1240 \\
& 3: GRPO (acc)          & .0614$^\dagger$ & .1397$^\dagger$ & .1233 \\
& 3: GRPO+Judge          & .0697 & .1395 & \textbf{.1255} \\
\bottomrule
\end{tabular}
\end{table}


Accuracy-only GRPO does not recover the degradation introduced by reasoning SFT. In contrast, GRPO+Judge substantially improves traditional offline recommendation effectiveness. For \sid{}, it achieves the best R@10 on two of the three domains, exceeding the no-reasoning baseline on Office ($.1646$ vs.\ $.1595$, $+$3.2\%) and Industrial ($.1401$ vs.\ $.1361$, $+$2.9\%). On Video Games, it recovers approximately 41\% of the gap ($.0705$ vs.\ $.0728$), although the no-reasoning baseline remains best. For \tit{}, GRPO+Judge achieves the best R@10 on Industrial ($.1255$ vs.\ $.1253$) and recovers approximately 45\% of the gap on Games ($.0697$ vs.\ $.0783$), but fails to recover the degradation on Office ($.1395$ vs.\ $.1587$). Overall, the judge reward benefits \sid{} more than \tit{}, with \sid{} achieving the best recommendation effectiveness on two domains while \tit{} does so on only one.

These results suggest that the degradation observed under the current reasoning formulation is not solely a consequence of introducing explicit reasoning, but is also influenced by the optimization objective. When reinforcement learning rewards both recommendation accuracy and reasoning behavior, much of the degradation can be recovered.

Finally, we ask whether the semantic understanding of an LLM judge is necessary or whether a cheaper dense reward is sufficient. We therefore replace the two judge components with a single embedding-similarity reward based on the cosine similarity between E5 embeddings~\cite{Wang2024E5} of the predicted and ground-truth items:
\[
R = 0.5 \cdot R_{\text{acc}} + 0.5 \cdot R_{\text{sim}}.
\]
This variant requires no LLM calls during training \revise{and, unlike GRPO+Judge, restricts policy gradients to item tokens only, since the embedding reward provides no signal for reasoning tokens}. As shown in Appendix~\ref{app:grpo-emb}, the embedding reward nearly matches the LLM judge on Office and Industrial for both representations but, like the judge, fails to recover the degradation on Games. \revise{Because GRPO+Emb changes only the reward density while retaining item-token-only gradients, and still nearly matches GRPO+Judge, reward density---not gradient flow through reasoning tokens or judge semantics---is the operative variable.}

\subsection{Reasoning Trace Quality}
\label{sec:trace-quality}


The previous section showed that introducing explicit reasoning does not consistently improve traditional offline recommendation effectiveness. We now ask whether improving the descriptive quality of the resulting reasoning traces explains this outcome.

We generate reasoning traces from the reasoning-SFT models using greedy decoding (max 320 tokens) on 100 test users per dataset (300 traces per representation). An LLM judge (Gemini 2.5 Pro) evaluates each trace together with the user's interaction history and recommended item across six quality dimensions, assigning a score from 1--5 with a short justification. To ensure the evaluation measures descriptive reasoning trace quality rather than token readability, all SID codes are decoded to their human-readable titles before judging. Recommendation relevance is evaluated separately by scoring the top-10 recommended items in human-readable form (Table~\ref{tab:trace-quality}, bottom).

\begin{table*}[t]
\caption{LLM judge scores (1--5) pooled over the three datasets ($n = 300$ traces per representation per stage). \emph{Top:} reasoning-trace quality across six dimensions. \emph{Bottom:} recommendation relevance (top-10 predicted items, scored in human-readable titles for all configurations). All SID traces are decoded---every SID code is replaced with its human-readable title before judging. \sid$^0$: lean single-task Stage~1; \sid{}-A$^1$: SIDReasoner's eight-task alignment Stage~1; \tit$^2$: title-based. SFT = Stage~2 reasoning model; +Judge = Stage~3 GRPO with LLM-judge reward. Higher is better. Superscripts indicate the model is significantly higher than the indexed model within the same stage ($p < 0.05$, one-sided Mann-Whitney $U$ with Bonferroni correction over 6 dimensions). Primary judge: Gemini 2.5 Pro.}
\label{tab:trace-quality}
\centering
\small
\begin{tabular}{lllllll}
\toprule
& \multicolumn{3}{c}{\textbf{Reasoning SFT}} & \multicolumn{3}{c}{\textbf{GRPO+Judge}} \\
\cmidrule(lr){2-4} \cmidrule(lr){5-7}
\textbf{Dimension} & \sid$^0$ & \sid{}-A$^1$ & \tit$^2$ & \sid$^0$ & \sid{}-A$^1$ & \tit$^2$ \\
\midrule
\multicolumn{7}{l}{\emph{Trace quality}} \\
Groundedness        & 1.58 & 3.97$^{0}$ & 4.07$^{0}$ & 1.43 & 3.99$^{0}$ & 4.24$^{0}$ \\
Specificity         & 4.02 & 4.65$^{0,2}$ & 4.41$^{0}$ & 4.10 & 4.66$^{0,2}$ & 4.44$^{0}$ \\
Logical Coherence   & 1.15 & 2.37$^{0,2}$ & 1.92$^{0}$ & 1.42 & 2.40$^{0,2}$ & 2.10$^{0}$ \\
Preference Extr.    & 1.48 & 3.77$^{0,2}$ & 3.36$^{0}$ & 1.61 & 3.76$^{0}$ & 3.56$^{0}$ \\
Rec.\ Justification & 1.08 & 1.39$^{0}$ & 1.29$^{0}$ & 1.15 & 1.33$^{0}$ & 1.19 \\
Interpretability    & 4.03 & 4.93$^{0}$ & 4.94$^{0}$ & 3.68 & 4.93$^{0}$ & 4.92$^{0}$ \\
\cmidrule{2-7}
Aggregate           & 2.22 & \textbf{3.51} & 3.33 & 2.23 & \textbf{3.51} & 3.41 \\
\midrule
\multicolumn{7}{l}{\emph{Recommendation quality}} \\
Rec.\ Relevance     & 3.64 & 3.71 & \textbf{3.88}$^{0}$ & \textbf{3.95}$^{1}$ & 3.64 & 3.85$^{1}$ \\
\bottomrule
\end{tabular}
\end{table*}

We use Gemini rather than GPT because the reasoning traces are generated by a GPT-distilled model, reducing the risk of circular evaluation. Each quality dimension is scored in a separate API call to avoid cross-dimension anchoring, and cross-family reliability is assessed with GPT-4o on a 15\% subset.

The six dimensions assess the descriptive quality of the generated reasoning traces: whether they accurately describe the user's history, inferred preferences, and recommendation rationale in a grounded and interpretable way. They include:
\begin{itemize}
    \item \textbf{Groundedness}: Are the reasoning claims supported by the user's history?
    \item \textbf{Specificity}: Does the trace reference concrete item attributes rather than generic statements?
    \item \textbf{Logical Coherence}: Does the reasoning form a consistent chain from history to recommendation?
    \item \textbf{Preference Extraction}: Does the trace correctly identify the user's preferences?
    \item \textbf{Recommendation Justification}: Does the trace explain why the recommended item follows from those preferences?
    \item \textbf{Interpretability}: Is the reasoning understandable and verifiable by a human reader?
\end{itemize}

Table~\ref{tab:trace-quality} reports the resulting scores for both the reasoning-SFT models (left columns) and the GRPO+Judge models (right columns).

\subsubsection{SFT traces.}
%
Title is judged significantly higher than the vanilla \sid{} model on all six trace-quality dimensions, even after decoding SID codes to human-readable titles. The largest difference is in \emph{groundedness} ($1.58 \rightarrow 4.07$): vanilla \sid{} traces frequently hallucinate items and preferences that are not supported by the user's interaction history. Appendix~D provides a representative qualitative example illustrating this behaviour.

Paying the full alignment tax closes this gap. The \sid{}-A model achieves the highest aggregate trace quality ($3.51$), exceeding Title ($3.33$), with significantly higher scores on specificity, logical coherence, and preference extraction. This confirms that the eight-task alignment successfully grounds semantic IDs in natural language.
Despite these improvements, both representations score near the floor on logical coherence and recommendation justification, suggesting that the distilled reasoning traces resemble plausible commentary rather than valid inferential chains.

\subsubsection{GRPO+Judge traces.}
%
In contrast to traditional offline recommendation effectiveness, descriptive reasoning trace quality changes very little under GRPO+Judge. Aggregate trace quality remains essentially unchanged for all representations: SID stays at $2.23$ (from $2.22$), \sid{}-A remains at $3.51$, and Title increases only slightly to $3.41$ (from $3.33$). The Title-over-SID gap is likewise preserved, with significant advantages on five of the six dimensions.

This stands in sharp contrast to the recommendation results in Table~\ref{tab:grpo-judge}. The judge reward improves \sid{} recommendation effectiveness beyond the no-reasoning baseline on two domains without improving its reasoning traces. Conversely, \sid{}-A achieves the highest reasoning quality ($3.51$) but the worst recommendation effectiveness. Together, these findings provide strong evidence that improvements in descriptive reasoning trace quality do not necessarily translate into improvements in traditional offline recommendation effectiveness.

\subsubsection{Recommendation relevance.}
%
The recommendation relevance row evaluates a different quantity from the trace-quality dimensions above. Rather than assessing the reasoning trace itself, it scores only the semantic relevance of the resulting recommendations after converting all outputs to human-readable titles, making the comparison independent of the underlying item representation.

For the SFT models, \tit{} recommendations are judged significantly more relevant than the vanilla \sid{} model ($3.88$ vs.\ $3.64$, $p<0.05$), while \sid{}-A falls between the two ($3.71$, not significantly different from either). GRPO+Judge substantially narrows this gap: vanilla \sid{} improves from $3.64$ to $3.95$, becoming the highest-scoring configuration, whereas \sid{}-A drops to $3.64$, significantly below both vanilla \sid{} ($p<0.01$) and \tit{} ($p<0.05$).

These recommendation relevance judgments exhibit a different pattern from descriptive reasoning trace quality. While GRPO+Judge substantially improves recommendation relevance for the vanilla \sid{} model, descriptive reasoning trace quality changes very little. Likewise, \tit{} produces substantially higher-quality descriptive reasoning traces than the vanilla \sid{} model, yet only modestly higher recommendation relevance under SFT.

Taken together with the offline results in Table \ref{tab:grpo-judge}, these findings suggest that descriptive reasoning trace quality, LLM-judged recommendation relevance, and traditional offline recommendation effectiveness capture different properties of recommendation systems. Improvements in descriptive reasoning traces do not consistently translate into improved traditional offline recommendation effectiveness, while LLM-judged recommendation relevance may exhibit different trends. Understanding the relationship between these evaluation perspectives remains an important direction for future work.


\subsection{Reasoning in a Production-Scale Setting}
\label{sec:spotify-playlist}

The controlled comparison above isolates the interaction between reasoning and item representation on academic benchmarks with small catalogs (${\sim}$3.5K items). A natural question is whether these findings extend to a production-scale system where the model must serve multiple tasks over a much larger catalog. Unlike the controlled Amazon experiments, this setting evaluates reasoning in a production-scale multi-task model with extensive catalog grounding, allowing us to examine whether the same relationship between descriptive reasoning traces and recommendation effectiveness persists at much larger scale.

We therefore evaluate reasoning over semantic IDs in a playlist continuation task, using a proprietary dataset of
$\mathcal{O}(10^6)$ playlists and a catalog of $\mathcal{O}(10^8)$ tracks.
Given a playlist title and its first $N$ tracks, the model generates likely
continuation tracks. For each training instance, a teacher language model
supplies a plausible, target-informed reasoning trace. For example:

\begin{quote}
  \small
  \noindent\texttt{<think>}\\
  \textbf{Clues:} this is a dreamy and calm playlist with indie and alternative
  styles; artists include the strokes and loathe.\\
  \textbf{Plan:} bridge into a related style; favor bedroom pop with a dreamy
  mood, prioritizing the strokes, lana del rey, and malcolm todd.\\
  \texttt{</think>}
\end{quote}

We use a concise two-line schema to provide consistent supervision.
\emph{Clues} summarizes input-visible mood, genre, and representative artists,
while \emph{Plan} specifies a continuation strategy followed by
target-informed style, mood, and artist cues.

\paragraph{Representation and training.}
Unlike the single-task Amazon models, this model is designed for a production setting where multiple tasks share a single backbone. The input interleaves textual
metadata with semantic identifier (\sid{}) tokens, reasoning is expressed in
natural language, and recommended tracks are generated as \sid{} tokens. Using the Qwen3
1.7B architecture, we first perform continued pretraining on a catalog-grounding objective that aligns SID and text tokens across multiple tasks (analogous to the eight-task alignment in Section~5.1.2), followed by two epochs of full-parameter
fine-tuning on the playlist continuation task with reasoning traces (equivalent to Stage~2 Reasoning SFT).

\paragraph{Evaluation.}
We compare No Reasoning and reasoning-conditioned candidate generation on the
same held-out set. No Reasoning begins directly with continuation tracks,
whereas the reasoning condition first generates a trace. As
Table~\ref{tab:spotify-results} shows, HR@30 and NDCG@30 remain broadly stable.
Reasoning nevertheless introduces suitable recommended tracks that are absent
from the No Reasoning output. Rank aggregation captures this complementary
signal and obtains the strongest value on both metrics. Positive and negative
examples are reported in Appendix~\ref{app:spotify-examples}.

\begin{table}[htbp]
  \centering
  \caption{Playlist-continuation quality on a fixed held-out panel. Rank
  aggregation takes the union of candidates from the No Reasoning and Reasoning
  outputs and scores each track by the sum of its reciprocal ranks.}
  \label{tab:spotify-results}
  \begin{tabular}{lcc}
    \toprule
    Method & HR@30 & NDCG@30 \\
    \midrule
    No Reasoning     & 0.6337 & 0.1364 \\
    Reasoning SFT       & 0.6343 & 0.1329 \\
    Rank aggregation & \textbf{0.6515} & \textbf{0.1380} \\
    \bottomrule
  \end{tabular}
\end{table}

These results extend the central observation from the controlled Amazon experiments to a production-scale multi-task setting: reasoning over (aligned) semantic IDs produces coherent, catalog-grounded traces that can steer recommendations, yet does not reliably improve aggregate recommendation effectiveness. 
Even in a production-scale setting with a multi-task model, a large catalog, and extensive catalog grounding, reasoning introduces useful candidates but displaces a comparable number. 
The rank aggregation result indicates that reasoning surfaces complementary candidates, and more selective teacher traces or tighter grounding between stated cues and continuation tracks may convert this complementary signal into consistent improvements. This suggests that reasoning changes which candidates are surfaced, rather than simply improving or degrading traditional offline recommendation effectiveness.
One possible explanation is the \sid{}-to-text-to-\sid{} round trip, in which catalog
evidence must be verbalized and then mapped back to identifiers. 
Additionally, our Amazon experiments showed that reinforcement learning with richer reward signals (GRPO+Judge) can partially recover the degradation introduced by reasoning SFT (Section~\ref{sec:grpo-judge}); applying similar RL with well-designed rewards to the playlist continuation task is a promising direction.


\section{Discussion}
\label{sec:discussion}

Taken together, the results suggest three broader implications for reasoning, alignment, and evaluation in generative recommendation.


\paragraph{Optimization objectives substantially influence the impact of explicit reasoning on traditional offline recommendation effectiveness.} Across all three domains, supervised reasoning and accuracy-only GRPO fail to improve traditional offline recommendation effectiveness: every reasoning delta is negative or zero, and accuracy-only GRPO remains close to its SFT warm-start. Our analysis attributes this failure to reward sparsity, with 70--96\% of prompts yielding no learning signal across all sampled generations. Replacing the sparse prefix-match reward with a composite LLM-judge reward substantially changes the picture. GRPO+Judge enables \sid{} reasoning to exceed the no-reasoning baseline on two of the three domains and recovers a substantial fraction of the degradation on Video Games for both representations. These results suggest that the degradation observed under the current reasoning formulation is not solely a consequence of introducing explicit reasoning, but is also influenced by the optimization objective and may also reflect limitations of single-ground-truth offline evaluation.
The production-scale playlist experiment (Section~\ref{sec:spotify-playlist}) provides further evidence: although reasoning SFT alone does not improve aggregate effectiveness, rank aggregation over reasoning and no-reasoning outputs yields the best performance on both metrics, indicating that reasoning surfaces complementary candidates that the no-reasoning model does not retrieve. Taken together, these results suggest that optimization objectives are a first-order factor in determining whether explicit reasoning translates into improved traditional offline recommendation effectiveness.


\paragraph{Descriptive reasoning trace quality, recommendation relevance, and traditional offline effectiveness capture different properties.}
The most consistent finding across our experiments is that descriptive reasoning trace quality, LLM-judged recommendation relevance, and traditional offline recommendation effectiveness need not move together. 
Natural-language titles substantially improve descriptive reasoning trace quality relative to the vanilla \sid{} model, yet do not consistently improve traditional offline recommendation effectiveness. Likewise, paying the full alignment tax produces the highest descriptive reasoning trace quality, but not the strongest recommendation effectiveness. Conversely, optimizing with an LLM-judge reward substantially improves recommendation relevance and partially recovers traditional offline recommendation effectiveness while producing little change in descriptive reasoning trace quality. Together, these results show that improving descriptive reasoning trace quality alone is not sufficient to consistently improve traditional offline recommendation effectiveness. Instead, descriptive reasoning trace quality, recommendation relevance, and traditional offline recommendation effectiveness appear to capture different properties of recommendation systems. One possible explanation for these differing behaviors is that reasoning and prediction interact differently depending on how item representations are encoded.

One possible explanation is representational interference. In the Title configuration, reasoning traces and prediction targets share the same natural-language vocabulary, so generating the reasoning trace may shift the output distribution toward plausible but incorrect items. By contrast, SID tokens occupy a disjoint vocabulary from the reasoning text, creating a natural separation between reasoning and prediction. 
This may explain why the LLM-judge reward benefits \sid{} more than \tit{}: improvements in descriptive reasoning traces may translate into better recommendations without directly interfering with item prediction.


Title SFT recommendations are nevertheless judged significantly more relevant by the LLM judge than those of the vanilla \sid{} model ($3.88$ vs.\ $3.64$, $p<0.05$), and SID GRPO+Judge achieves the highest relevance score ($3.95$) across all configurations.
The differing behavior of LLM-judged recommendation relevance and traditional offline recommendation effectiveness further supports the view that these evaluation perspectives capture different properties.
Recent work suggests that conventional train--test splits provide severely incomplete relevance labels, producing unstable model rankings~\cite{Penha2025LLMJudge}. If similar incompleteness affects our Amazon benchmarks, the observed recommendation penalty may overstate the true cost of reasoning. Whether traditional effectiveness metrics or LLM-judge relevance better reflects user utility remains an open question that ultimately requires online evaluation.

\revise{\paragraph{Answer-conditioned teacher traces as a candidate root cause.} A complementary explanation is a train/test mismatch in the teacher traces. The Stage~2 traces are \emph{answer-conditioned}: a GPT model receives both the user's history and the known target item, constructing a preference narrative steered by knowledge of the answer. At test time, the model must generate a trace \emph{before} knowing the target. The student learns to produce fluent preference analyses but lacks the teacher's access to the target that originally guided them. This is consistent with the low logical coherence (1.15--2.40) and recommendation justification (1.08--1.39) scores in Table~\ref{tab:trace-quality}: traces describe preferences fluently but do not form inferential chains that reliably point to specific items. The disconnect may thus partly reflect a limitation of answer-conditioned distillation rather than an inherent property of reasoning in generative recommendation. End-to-end approaches~\cite{Kong2025RecZero} may behave differently.}

\revise{\paragraph{A note on terminology.} Throughout this paper, we follow prevailing terminology and refer to intermediate tokens as ``reasoning traces.'' Kambhampati et al.~\cite{Kambhampati2025StopAnthropomorphizing} argue that this anthropomorphizes what may be better understood as learned prompt augmentations---intermediate tokens whose task utility is independent of their semantic content. Our results support this view: reward density, not trace semantics, drives the GRPO recovery (Appendix~\ref{app:grpo-emb}), and the decoupling between trace quality and recommendation effectiveness is what one would expect under the prompt-augmentation framing. We retain the term for consistency with the literature but caution against treating descriptive trace quality as a proxy for functional utility.}


\paragraph{The alignment tax primarily affects descriptive reasoning traces.} One might expect semantic IDs to suffer a larger recommendation penalty because of the additional alignment required for reasoning. Our results suggest a different picture. The vanilla \sid{} model produces lower-quality descriptive reasoning traces than the \tit{} model, but incurs a smaller recommendation penalty. Paying the full alignment tax reverses this pattern: \sid{}-A achieves the highest descriptive reasoning trace quality in Table~\ref{tab:trace-quality}, yet the worst recommendation effectiveness in Table~\ref{tab:grpo-judge}, with GRPO further degrading an already weakened reasoning model. 
These results suggest that the alignment tax is real, but it manifests primarily in descriptive reasoning trace quality rather than traditional offline recommendation effectiveness. Extensive alignment successfully teaches the model to produce substantially higher-quality descriptive reasoning traces over semantic IDs, yet these improvements do not translate into better traditional offline recommendation effectiveness and may even move the model away from the calibrated next-item distribution learned during no-reasoning training.


\paragraph{Limitations.} This study has several limitations. First, we evaluate a single backbone (Qwen3-1.7B); the relationship between descriptive reasoning trace quality and traditional offline recommendation effectiveness may differ for larger models with greater capacity. Second, Stage~2 is initialized from GPT-generated teacher traces rather than the model's own reasoning. Although Stage~3 optimizes from self-generated traces, the supervised initialization may constrain the reasoning style. End-to-end approaches such as OneRec-Think~\cite{Liu2025OneRecThink}, which develop reasoning without teacher distillation, may therefore behave differently. Finally, our evaluation is entirely offline. 
Whether the observed relationships among descriptive reasoning trace quality, recommendation relevance, and traditional offline recommendation effectiveness translate into user satisfaction remains an open question requiring online evaluation.


\section{Conclusion}


We presented the first controlled comparison of explicit descriptive reasoning traces across semantic IDs and natural-language titles in generative recommendation. Across both representations, standard reasoning training degrades traditional offline recommendation effectiveness, while a richer LLM-judge reward partially recovers this degradation. Paying the full alignment tax substantially improves descriptive reasoning trace quality but does not improve traditional offline recommendation effectiveness.

Our central finding is that improving descriptive reasoning trace quality alone is not sufficient to consistently improve traditional offline recommendation effectiveness under the training objectives and evaluation protocols studied here. 
Across multiple independent interventions --- including natural-language titles, extensive SID alignment, and richer optimization objectives --- we observed that descriptive reasoning trace quality, recommendation relevance, and traditional offline recommendation effectiveness need not move together, suggesting that they capture different aspects of recommender performance. Our results suggest that item representation, optimization objectives, and evaluation protocols jointly influence how improvements in descriptive reasoning traces translate into recommendation behavior. We hope these findings motivate future work on optimization objectives and evaluation protocols that better translate improvements in descriptive reasoning traces into improved recommendation behavior.

\section*{Declaration on Generative AI}
During the preparation of this work, the authors used Claude Code for code assistance, for running experiments and formatting. The authors reviewed and edited the content as needed and takes full responsibility for the publication's content.

\bibliography{references}

\appendix

\section{LLM Judge Prompts}
\label{app:judge-prompts}

\paragraph{System prompt (trace quality).}
\begin{quote}\small\ttfamily
You are a careful evaluator of recommendation-system reasoning traces. Score ONLY the dimension asked, on an integer 1--5 scale, using the rubric. Reply with JSON: \{"score": <int 1-5>, "justification": "<one sentence>"\}.
\end{quote}

\paragraph{User prompt (trace quality).}
Each dimension is scored in a separate API call. SID traces are decoded---every SID code is replaced with its human-readable title---before being passed to the judge.
\begin{quote}\small\ttfamily
Dimension: \{label\}\\
Rubric: \{rubric\}\\[4pt]
User interaction history (in the model's item representation):\\
\{history\}\\[4pt]
Recommended next item: \{item\}\\[4pt]
Reasoning trace to evaluate:\\
\{trace\}\\[4pt]
Score \{label\} 1--5 with the rubric.
\end{quote}

\paragraph{Rubrics.}
\begin{itemize}\small
    \item \textbf{Groundedness}: Does the trace reference items/attributes actually present in the user's history? 1 = fabricates items/interactions not in the history; 5 = every claim is traceable to a specific history item.
    \item \textbf{Specificity}: Does the trace mention concrete product attributes (genre, brand, material, price range, use-case) vs.\ vague statements like ``the user likes similar items''? 1 = entirely generic; 5 = references multiple concrete attributes.
    \item \textbf{Logical Coherence}: Does the reasoning chain follow logically from observed history $\to$ inferred preferences $\to$ recommendation? 1 = non-sequiturs, circular reasoning, or contradictions; 5 = clear valid inferential chain.
    \item \textbf{Preference Extraction}: Does the trace correctly identify the user's preferences from their history? 1 = no/incorrect preferences; 5 = nuanced, accurate patterns including temporal trends.
    \item \textbf{Recommendation Justification}: Does the trace explain why this specific item is the right next recommendation, connecting it to extracted preferences? 1 = no connection; 5 = the recommendation follows naturally and specifically from the reasoning.
    \item \textbf{Interpretability}: Can a human understand what items/attributes the trace refers to? 1 = references are opaque or meaningless; 5 = fully human-readable and understandable.
\end{itemize}

\paragraph{System prompt (recommendation relevance).}
\begin{quote}\small\ttfamily
You are an expert recommendation evaluator. Score ONLY the dimension asked, on an integer 1--5 scale, using the rubric. Reply with JSON: \{"score": <int 1-5>, "justification": "<one sentence>"\}.
\end{quote}

\paragraph{User prompt (recommendation relevance).}
Both configurations' histories and predictions are shown in human-readable titles.
\begin{quote}\small\ttfamily
Dimension: Recommendation Relevance\\
Rubric: Based on the user's purchase/interaction history, how relevant are the recommended items as their next purchase? Consider category/use-case fit, complementarity, and progression of intent. 1 = irrelevant; 2 = weakly related; 3 = plausible; 4 = relevant; 5 = highly relevant. You are scoring the RECOMMENDATIONS themselves, not a reasoning trace.\\[4pt]
User interaction history (item titles, oldest to newest):\\
- \{title\_1\}\\
- \{title\_2\}\\
\ldots\\[4pt]
Recommended next items (ranked):\\
1. \{pred\_title\_1\}\\
2. \{pred\_title\_2\}\\
\ldots\\[4pt]
Score Recommendation Relevance 1--5 with the rubric.
\end{quote}

\section{GRPO LLM-Judge Reward Prompt}
\label{app:rl-reward-prompt}

The composite reward used in GRPO+Judge (Section~\ref{sec:grpo-judge}) scores each sampled generation with a single LLM call that returns two 1--5 scores. The judge always receives title-based representations regardless of the model's native item representation.

\paragraph{System prompt.}
\begin{quote}\small\ttfamily
You are a recommendation reasoning evaluator. You receive a user's purchase history (item titles), a reasoning trace from a recommender model, and the recommended item title.\\[4pt]
Score two dimensions (1--5 integers):\\[4pt]
1. \textbf{trace\_quality}: Is the reasoning coherent, specific, and well-structured? Does it identify concrete user preferences and build a logical chain from history to recommendation? (1=incoherent/generic template, 5=specific, logical, well-grounded)\\[4pt]
2. \textbf{relevance\_match}: Based on the user's history, how relevant is the recommended item? Does it match the patterns and preferences visible in the history? (1=completely unrelated, 5=highly relevant continuation)\\[4pt]
Respond with JSON only: \{"trace\_quality": <int>, "relevance\_match": <int>\}
\end{quote}

\paragraph{User prompt.}
\begin{quote}\small\ttfamily
History: \{history\_titles\}\\[4pt]
Reasoning trace: \{trace\}\\[4pt]
Recommended item: \{predicted\_title\}
\end{quote}

Both scores are normalized from 1--5 to $[0, 1]$ via $(s - 1) / 4$, then combined with the prefix-match accuracy reward: $R = 0.5 \cdot R_{\text{acc}} + 0.25 \cdot R_{\text{trace}} + 0.25 \cdot R_{\text{rel}}$. The judge model is GPT-4o-mini with temperature~0.

\section{GRPO with Embedding Similarity Reward}
\label{app:grpo-emb}

As a cheaper alternative to the LLM-judge reward, we combine the prefix-match accuracy reward with a dense embedding-similarity signal:
\begin{equation}
    R = 0.5 \cdot R_{\text{acc}} + 0.5 \cdot R_{\text{sim}}
\end{equation}
where $R_{\text{sim}} = \max(0,\; \cos(\mathbf{e}_{\text{pred}}, \mathbf{e}_{\text{target}}))$ is the cosine similarity between pre-computed E5 embeddings~\cite{Wang2024E5} of the predicted and ground-truth items.
Unlike the judge reward, this variant uses item-focused GRPO (gradient restricted to item tokens after \texttt{</think>}), since the embedding reward depends only on the predicted item.

\begin{table}[h]
\caption{R@10 across RL reward variants. GRPO (acc): prefix-match reward only; GRPO+Judge: composite LLM-judge reward (Section~\ref{sec:grpo-judge}); GRPO+Emb: embedding-similarity reward. Best reasoning configuration per domain in \textbf{bold}.}
\label{tab:grpo-emb}
\centering
\small
\begin{tabular}{llccc}
\toprule
\textbf{Repr.} & \textbf{Stage} & \textbf{Games} & \textbf{Office} & \textbf{Industrial} \\
\midrule
\multirow{5}{*}{\sid}
& 1: No reasoning       & .0728 & .1595 & .1361 \\
& 2: Reasoning SFT      & .0689 & .1589 & .1324 \\
& 3: GRPO (acc)          & .0682 & .1591 & .1332 \\
& 3: GRPO+Judge          & \textbf{.0705} & \textbf{.1646} & \textbf{.1401} \\
& 3: GRPO+Emb            & .0682 & .1640 & .1399 \\
\midrule
\multirow{5}{*}{\tit}
& 1: No reasoning       & .0783 & .1587 & .1253 \\
& 2: Reasoning SFT      & .0627 & .1416 & .1240 \\
& 3: GRPO (acc)          & .0614 & .1397 & .1233 \\
& 3: GRPO+Judge          & \textbf{.0697} & .1395 & .1255 \\
& 3: GRPO+Emb            & .0615 & \textbf{.1478} & \textbf{.1273} \\
\bottomrule
\end{tabular}
\end{table}

The embedding reward nearly matches the judge for \sid{} on Office ($.1640$ vs.\ $.1646$) and Industrial ($.1399$ vs.\ $.1401$), and \emph{exceeds} the judge for \tit{} on Office ($.1478$ vs.\ $.1395$) and Industrial ($.1273$ vs.\ $.1255$).
On Games---the domain where all reward variants struggle---the embedding reward matches accuracy-only GRPO for both representations.
These results suggest that reward density is more important than semantic richness: a simple cosine similarity provides a comparable training signal to an LLM judge at zero API cost.

\section{Amazon Qualitative Example}
\label{app:amazon-example}

Figure~\ref{fig:trace-example} shows a representative case from the Video Games dataset.
The user's history is entirely retro gaming: an N64 console, GoldenEye 007, a PS1 system with memory card, Super Mario Bros.\ 3, and a SNES extension cable.
The Title trace correctly identifies this as a \emph{retro and classic gaming} pattern: ``classic gaming experiences and nostalgia for iconic titles like Super Mario Bros.\ 3 and GoldenEye 007,'' and recommends retro accessories that match.
The SID trace, unable to read its own opaque tokens, defaults to the most common template in its training data: ``a strong inclination towards action-adventure and RPG genres''---then recommends \texttt{<a\_216><b\_54><c\_190>} and \texttt{<a\_201><b\_145><c\_9>} (Zelda: Breath of the Wild and The Witcher 3, respectively), modern open-world games that completely miss the retro pattern.
The SID trace scores $1$ on groundedness and preference extraction; the Title trace scores $5$ on both.
This pattern---SID traces producing plausible-sounding but unverifiable generic narratives---recurs across the dataset.

\begin{figure}[t]
\centering
\small
\fbox{\parbox{0.95\columnwidth}{
\textbf{User history:} Nintendo 64 System $\mid$ GoldenEye 007 $\mid$ Sony PlayStation 1 System $\mid$ PS1 Memory Card $\mid$ Super Mario Bros.\ 3 $\mid$ Retro-Bit SNES Extension Cable\\[4pt]
\textbf{Target:} Sonic the Hedgehog\\[6pt]
\textbf{\sid{} trace:} ``Based on your strong inclination towards \colorbox{red!15}{\sout{action-adventure and RPG genres}}, I recommend exploring titles like \colorbox{red!15}{\sout{\texttt{<a\_216><b\_54><c\_190>}}} for its \colorbox{red!15}{\sout{immersive world and character development}}. Additionally, consider games that blend action with narrative depth, such as \colorbox{red!15}{\sout{\texttt{<a\_201><b\_145><c\_9>}}}\ldots''\\[4pt]
\textbf{\tit{} trace:} ``Based on your strong inclination towards \colorbox{teal!15}{\underline{classic gaming experiences}} and \colorbox{teal!15}{\underline{nostalgia for iconic titles like Super Mario Bros.\ 3 and GoldenEye 007}}, I recommend exploring more \colorbox{teal!15}{\underline{retro consoles or accessories}} that enhance your gaming setup. Consider investing in a high-quality controller\ldots look for \colorbox{teal!15}{\underline{accessories that improve your gaming experience, such as extension cables or memory cards}}\ldots''
}}
\caption{Reasoning traces for the same user, generated by the \sid{} and \tit{} reasoning-SFT models. The user's history is entirely retro gaming, yet the SID trace hallucinates ``action-adventure and RPG genres'' and recommends modern open-world titles (\texttt{<a\_216><b\_54><c\_190>} = Zelda: Breath of the Wild; \texttt{<a\_201><b\_145><c\_9>} = The Witcher 3---titles shown here for readability only; the actual trace contains only opaque SID codes). The Title trace correctly identifies the retro/nostalgia pattern. \colorbox{red!15}{\sout{Strikethrough/red}}: hallucinated preferences; \colorbox{teal!15}{\underline{underline/teal}}: grounded, correct preferences.}
\label{fig:trace-example}
\end{figure}

Crucially, higher descriptive reasoning trace quality does \emph{not} consistently translate into improved traditional offline recommendation effectiveness.
This relationship appears in multiple forms throughout our experiments: Title SFT produces higher-quality traces ($3.33$) yet suffers a larger ranking penalty than SID ($-$23\% vs.\ $-$11\% N@10 on Games); \sid{}-A achieves the \emph{highest} trace quality ($3.51$) yet the \emph{worst} recommendations (Table~\ref{tab:grpo-judge}); and SID GRPO+Judge improves recommendations to \emph{exceed} the baseline on two domains while its trace quality remains unchanged ($2.23$).
While Title SFT recommendations are judged more relevant by the LLM judge ($3.88$ vs $3.64$, $p < 0.05$), this advantage does not offset the ranking penalty under traditional offline metrics.

\section{Playlist Continuation Qualitative Examples}
\label{app:spotify-examples}

The following examples show how generated traces steer the continuation. The
positive case identifies the dominant continuation artist, whereas the negative
case follows the wrong branch of a mixed playlist.

\begin{center}
\begin{minipage}{\linewidth}
  \centering
  \captionof{table}{Positive example. Green marks held-out continuation tracks
  retrieved after generating the reasoning trace.}
  \label{tab:spotify-positive}
  \scriptsize
  \setlength{\tabcolsep}{3.5pt}
  \renewcommand{\arraystretch}{1.08}
  \begin{tabular}{@{}p{0.10\linewidth}p{0.42\linewidth}p{0.42\linewidth}@{}}
    \toprule
    \multicolumn{3}{@{}l}{\textbf{Playlist 1}} \\
    \midrule
    \textbf{Input}
      & \multicolumn{2}{p{0.84\linewidth}@{}}{
        Smells Like Teen Spirit --- Nirvana; I Was Made For Lovin' You --- KISS;
        Paranoid - 2009 Remaster --- Black Sabbath; You're Gonna Go Far, Kid ---
        The Offspring; Nookie --- Limp Bizkit; B.Y.O.B. --- System Of A Down;
        Forty Six \& 2 --- TOOL; Awake --- Godsmack
        \emph{(8 of 12 tracks shown)}
      } \\
    \textbf{Targets}
      & \multicolumn{2}{p{0.84\linewidth}@{}}{Hated --- YUNGBLUD; 18 ---
        Anarbor; Dirty Little Secret --- The All-American Rejects; Cherry Waves
        --- Deftones; Kilby Girl --- The Backseat Lovers; My Own Summer (Shove
        It) --- Deftones; Change (In the House of Flies) --- Deftones; Sextape
        --- Deftones; Entombed --- Deftones \emph{(9 resolved targets shown)}} \\
    \midrule
    \textbf{Reasoning}
      & \multicolumn{2}{p{0.84\linewidth}@{}}{
        \texttt{<think>}\newline
        \textbf{clues:} this is an aggressive and dark playlist with rock and
        metal styles; artists include nirvana and korn.\newline
        \textbf{plan:} stay close to the playlist's core; favor nu metal with an
        intense mood, prioritizing nirvana, linkin park, and deftones.\newline
        \texttt{</think>}
      } \\
    \midrule
    \textbf{Rank} & \textbf{No Reasoning} & \textbf{Reasoning} \\
    \midrule
    1  & Papercut --- Linkin Park
       & \gain{My Own Summer (Shove It) --- Deftones} \\
    2  & Faint --- Linkin Park
       & \gain{Change (In the House of Flies) --- Deftones} \\
    3  & Given Up --- Linkin Park
       & Faint --- Linkin Park \\
    4  & Can You Feel My Heart --- Bring Me The Horizon
       & Heart-Shaped Box --- Nirvana \\
    5  & Bodies --- Drowning Pool
       & Papercut --- Linkin Park \\
    6  & Psychosocial --- Slipknot
       & Come As You Are - Remastered --- Nirvana \\
    7  & Before I Forget --- Slipknot
       & Given Up --- Linkin Park \\
    8  & The Devil in I --- Slipknot
       & Be Quiet and Drive (Far Away) --- Deftones \\
    9  & Break Stuff --- Limp Bizkit
       & Somewhere I Belong --- Linkin Park \\
    10 & Somewhere I Belong --- Linkin Park
       & Blind --- Korn \\
    \bottomrule
  \end{tabular}
\end{minipage}
\end{center}

The input supports several prominent nu-metal bands, but the targets  leans toward Deftones. The explicit Deftones cue moves \emph{My Own
Summer (Shove It)} and \emph{Change (In the House of Flies)} to ranks 1 and 2.

\begin{center}
\begin{minipage}{\linewidth}
  \centering
  \captionof{table}{Negative example. Red marks a held-out continuation track
  retrieved without reasoning but displaced after following the trace.}
  \label{tab:spotify-negative}
  \scriptsize
  \setlength{\tabcolsep}{3.5pt}
  \renewcommand{\arraystretch}{1.08}
  \begin{tabular}{@{}p{0.10\linewidth}p{0.42\linewidth}p{0.42\linewidth}@{}}
    \toprule
    \multicolumn{3}{@{}l}{\textbf{Playlist  2} } \\
    \midrule
    \textbf{Input}
      & \multicolumn{2}{p{0.84\linewidth}@{}}{
        Maria I'm Drunk (feat. Justin Bieber \& Young Thug) --- Travis Scott;
        Heads Will Roll --- Yeah Yeah Yeahs; Damn I Love Miami --- Pitbull; party
        4 u --- Charli xcx; Uber Everywhere (feat. Travis Scott) --- MadeinTYO;
        MY HUMPS JUST WANNA VROOM VROOM --- dj g2g; Fireball (feat. John Ryan) ---
        Pitbull; P*RNSTAR --- Nessa Barrett
        \emph{(8 of 12 tracks shown)}
      } \\
    \textbf{Targets}
      & \multicolumn{2}{p{0.84\linewidth}@{}}{Dangerous Woman --- Ariana Grande;
        Everyday --- Ariana Grande; CRANK --- Slayyyter; My Humps --- Black Eyed
        Peas; If U Seek Amy --- Britney Spears; Where Have You Been --- Rihanna;
        Rush --- Troye Sivan; Pour It Up --- Rihanna; Circus --- Britney Spears;
        365 --- Charli xcx; Luxury --- Azealia Banks; Yummy --- Ayesha Erotica
        \emph{(12 of 16 targets shown)}} \\
    \midrule
    \textbf{Reasoning}
      & \multicolumn{2}{p{0.84\linewidth}@{}}{
        \texttt{<think>}\newline
        \textbf{clues:} this is an energetic playlist with a pop style; artists
        include travis scott and yeah yeah yeahs.\newline
        \textbf{plan:} bridge into a related style; favor american hip hop with
        an energetic mood, prioritizing travis scott, a\$ap rocky, and kanye
        west.\newline
        \texttt{</think>}
      } \\
    \midrule
    \textbf{Rank} & \textbf{No Reasoning} & \textbf{Reasoning} \\
    \midrule
    1  & Blow --- Kesha
       & Fashion Killa --- A\$AP Rocky \\
    2  & Promiscuous --- Nelly Furtado
       & CARNIVAL --- ¥\$ \\
    3  & 23 --- Mike WiLL Made-It
       & Nightcrawler (feat. Swae Lee \& Chief Keef) --- Travis Scott \\
    4  & Von dutch --- Charli xcx
       & Lord Pretty Flacko Jodye 2 (LPFJ2) --- A\$AP Rocky \\
    5  & Gimme More --- Britney Spears
       & HELICOPTER --- A\$AP Rocky \\
    6  & S\&M --- Rihanna
       & Ni**as In Paris --- JAY-Z \\
    7  & I Love It (feat. Charli XCX) --- Icona Pop
       & 90210 (feat. Kacy Hill) --- Travis Scott \\
    8  & The Way I Are --- Timbaland
       & FE!N (feat. Playboi Carti) --- Travis Scott \\
    9  & \loss{365 --- Charli xcx}
       & Peso --- A\$AP Rocky \\
    10 & Bad Girls --- M.I.A.
       & Mercy --- Kanye West \\
    \bottomrule
  \end{tabular}
\end{minipage}
\end{center}

The input mixes rap and dance-pop, while the targets mostly
follows the pop branch. The plan instead prioritizes Travis Scott, A\$AP Rocky,
and Kanye West; the model follows these cues, producing an exclusively rap
continuation and displacing the relevant pop candidates.

{\color{revision}
\section{BM25 Resolution Analysis}
\label{app:bm25-resolution}

Because the \tit{} configuration generates unconstrained text that is post-hoc resolved to catalog items via BM25, the larger reasoning penalty observed for \tit{} (Table~\ref{tab:main}) could partly reflect reasoning-induced drift in generated titles that degrades BM25 matching accuracy, rather than changes in the model's underlying item preferences. To disentangle these effects, we analyze the raw (pre-resolution) generations of the no-reasoning and reasoning \tit{} models on the full test sets, replicating the evaluation protocol of Section~4.

\paragraph{Faithful-generation rate.}
Because the item phase is capped at 32 tokens, titles longer than the budget can never be generated verbatim (28\% of Office titles exceed it), so a raw verbatim match rate mostly measures title length. We therefore measure the \emph{prefix-faithful} rate: the fraction of generations that verbatim match a catalog title, or a truncated prefix of one, after the normalization used by the BM25 index. Such generations are faithful catalog-title text; the only resolution work left is prefix completion, which affects both arms identically. Table~\ref{tab:resolution} reports both rates. Reasoning does not reduce the prefix-faithful rate on any domain: Office is identical ($.980$ vs.\ $.980$ at rank~1), and Video Games and Industrial are slightly \emph{higher} with reasoning. The premise of the confound---that reasoning degrades the resolvability of generated titles---does not hold.

\begin{table}[h]
\caption{Raw \tit{} generations against the catalog on the full test sets. Verbatim: exact match. Prefix-faithful: exact match or truncated verbatim prefix of a catalog title (the item phase is capped at 32 tokens, so long titles cannot appear verbatim). Rank-1: top beam; all beams: mean over the 10 evaluation beams.}
\label{tab:resolution}
\centering
\small
\setlength{\tabcolsep}{4pt}
\begin{tabular}{lcccccc}
\toprule
& \multicolumn{2}{c}{\textbf{Verbatim (rank-1)}} & \multicolumn{2}{c}{\textbf{Pref.-faithful (rank-1)}} & \multicolumn{2}{c}{\textbf{Pref.-faithful (all)}} \\
\cmidrule(lr){2-3} \cmidrule(lr){4-5} \cmidrule(lr){6-7}
\textbf{Domain} & No Reas. & Reas. & No Reas. & Reas. & No Reas. & Reas. \\
\midrule
Video Games        & .966 & .962 & .981 & .984 & .928 & .926 \\
Office Products    & .541 & .524 & .980 & .980 & .889 & .862 \\
Industrial \& Sci. & .427 & .419 & .960 & .975 & .674 & .736 \\
\bottomrule
\end{tabular}
\end{table}

\paragraph{Oracle resolution bound.}
To bound what any better resolver could contribute, we re-score the same generations with an \emph{oracle} resolver that maps any drifted (non-catalog) generation to the target item whenever the E5 embedding similarity between the generated string and the target title exceeds a threshold. Because the oracle peeks at the target, its scores upper-bound every deployable resolver, including dense retrieval. Table~\ref{tab:oracle} shows Recall@10. Applying the oracle to \emph{both} arms leaves the reasoning gaps intact on Video Games and Office at every threshold; on Industrial---where the Table~\ref{tab:main} reasoning delta is not statistically significant---the oracle in fact slightly favors the reasoning arm. (The $\geq 0.90$ threshold is loose for Industrial's near-duplicate catalog, where it conflates product variants such as filament colors; $\geq 0.95$ is the meaningful bound.) No resolver improvement can close the reasoning deficit.

\begin{table}[h]
\caption{Recall@10 on the full test sets under the actual BM25 resolution and under oracle resolvers (E5 cosine to the target $\geq 0.95$ / $\geq 0.90$) that upper-bound any deployable resolver.}
\label{tab:oracle}
\centering
\small
\begin{tabular}{llccc}
\toprule
\textbf{Domain} & \textbf{Arm} & \textbf{BM25} & \textbf{Oracle$_{.95}$} & \textbf{Oracle$_{.90}$} \\
\midrule
\multirow{2}{*}{Video Games} & No Reasoning & .079 & .079 & .087 \\
                             & Reasoning    & .063 & .064 & .071 \\
\midrule
\multirow{2}{*}{Office}      & No Reasoning & .166 & .175 & .198 \\
                             & Reasoning    & .149 & .155 & .176 \\
\midrule
\multirow{2}{*}{Industrial}  & No Reasoning & .111 & .187 & .268 \\
                             & Reasoning    & .120 & .196 & .246 \\
\bottomrule
\end{tabular}
\end{table}

\paragraph{Categorizing the drifted generations.}
We define a generation as \emph{drifted} when the rank-1 beam's text, after lowercasing and whitespace stripping (i.e., the same normalization used by the BM25 index), does not exactly match any catalog title. For non-drifted generations, BM25 resolution is exact and no confound is possible; drifted generations are the only cases where BM25's fuzzy matching could introduce errors. To understand what these drifted generations contain, we classify each into one of six categories: \emph{trace leak} (reasoning-style prose rather than a title), \emph{history echo} (a near-copy of an item already in the user's history), \emph{other-item near-miss} (a slightly misspelled version of a real but different catalog item, which BM25 resolves safely), \emph{target near-miss} (a reworded version of the target---the only category where a resolution confound could reside), \emph{hallucinated product} (a plausible but nonexistent product variant), and \emph{degenerate} (empty output). Classification uses lexical (difflib $\geq 0.80$) and semantic (E5 $\geq 0.95$) similarity against the target, the history, and the top BM25 candidates. Table~\ref{tab:drift-categories} reports the breakdown.

\begin{table}[h]
\caption{Breakdown of drifted (not prefix-faithful) rank-1 generations by category on the full test sets. Target near-miss is the only category where a resolution confound could reside. ``Other'' groups non-title text and degenerate (empty) outputs.}
\label{tab:drift-categories}
\centering
\small
\setlength{\tabcolsep}{4pt}
\begin{tabular}{llccccccc}
\toprule
\textbf{Domain} & \textbf{Arm} & \textbf{Drifted} & \textbf{Trace leak} & \textbf{Hist.\ echo} & \textbf{Other item} & \textbf{Target near} & \textbf{Halluc.} & \textbf{Other} \\
\midrule
\multirow{2}{*}{Video Games} & No Reasoning & 118 (1.9\%) & 0  & 69 & 29 & 2  & 8  & 10 \\
                             & Reasoning    & 99 (1.6\%)  & 4  & 46 & 40 & 3  & 6  & 0 \\
\midrule
\multirow{2}{*}{Office}      & No Reasoning & 97 (2.0\%)  & 0  & 56 & 9  & 12 & 18 & 2 \\
                             & Reasoning    & 97 (2.0\%)  & 3  & 57 & 16 & 7  & 13 & 1 \\
\midrule
\multirow{2}{*}{Industrial}  & No Reasoning & 181 (4.0\%) & 0  & 68 & 107 & 6 & 0  & 0 \\
                             & Reasoning    & 114 (2.5\%) & 10 & 66 & 12 & 25 & 1  & 0 \\
\bottomrule
\end{tabular}
\end{table}

Drift volume is symmetric between arms (Office: 97 vs.\ 97) or lower for the reasoning arm (Industrial: 114 vs.\ 181), and is dominated by history echoes and wrong-item near-misses in both arms. \emph{Target near-misses}---the only category where a resolution confound could reside---remain rare overall: at most 25 of 4{,}533 instances (0.55\%) on Industrial reasoning. On Industrial the reasoning arm does produce more target near-misses than the no-reasoning arm (25 vs.\ 6), but the oracle bound in Table~\ref{tab:oracle} shows that even resolving \emph{all} drifted generations to the target recovers recall symmetrically for both arms (${\sim}.187$--$.189$), so this difference does not bias the comparison. On Video Games and Office, target near-misses are negligible and balanced across arms. Trace leakage---reasoning prose spilling into the item phase---is negligible (4, 3, and 10 cases on Video Games, Office, and Industrial), consistent with the reasoning traces terminating within budget: at most $2.9\%$ of traces hit the 512-token cap on any domain.

\paragraph{Conclusion.}
The decoding asymmetry between the \sid{} and \tit{} arms cannot explain the \tit{} reasoning deficit in Table~\ref{tab:main}: reasoning does not reduce the rate of faithful, resolvable title generation, the drifted residue is small and symmetric across arms, and the oracle bound shows that even perfect resolution would not close the gap. The deficit therefore reflects changes in the model's item predictions, consistent with the preference-level explanations discussed in Section~\ref{sec:discussion}.
}

\end{document}